\documentclass[aps,prb,nofootinbib,twocolumn,superscriptaddress]{revtex4}
\usepackage{graphicx}
\newcommand{\be}{\begin{eqnarray}}
\newcommand{\ee}{\end{eqnarray}}
\newcommand{\la}{\langle}

\newcommand{\ds}[1]{#1{\hskip-2.2mm}/}

\newcommand{\hf}{\displaystyle{\not}}

\begin{document}

\title{The Effect of  Interactions on the Conductance of Graphene Nanoribbons}
\author{M.~Bazzanella}
\email{matteo.bazzanella@gmail.com}
\affiliation{Dipartimento di Fisica  Universit\'a degli Studi di Trento, Via Sommarive 14, Povo (Trento), I-38050 Italy.}
\author{P.~Faccioli}
\email{faccioli@science.unitn.it}
\affiliation{Dipartimento di Fisica  Universit\'a degli Studi di Trento, Via Sommarive 14, Povo (Trento), I-38050 Italy.}
\affiliation{I.N.F.N, Gruppo Collegato di Trento, Via Sommarive 14, Povo (Trento), I-38050 Italy.}
\author{ E.~Lipparini}
\email{lipparin@science.unitn.it}
\affiliation{Dipartimento di Fisica  Universit\'a degli Studi di Trento, Via Sommarive 14, Povo (Trento), I-38050 Italy.}
\affiliation{I.N.F.N, Gruppo Collegato di Trento, Via Sommarive 14, Povo (Trento), I-38050 Italy.}

\begin{abstract}
We study the effects of the interaction between electrons and holes on the conductance $G$
of quasi-one-dimensional graphene systems. 
 We 
first consider as a benchmark the limit in which all interactions are negligible, recovering the predictions of the tight-binding approximation for the spectrum of the system, and the well-known result $G=4~e^2/h$ for the lowest conductance quantum. Then we consider an exactly solvable field theoretical model in which the electro-magnetic interactions are  effectively local. Finally, we use the effective field theory formalism to develop an exactly solvable model in which we also include the effect of non-local interactions. We find that such interactions turn the nominally metallic armchair graphene nanoribbon into a semi-conductor,  while the short-range interactions lead to a correction to the  $G=4e^2/h$ formula. 
\end{abstract}
\maketitle

\flushbottom
\maketitle
\section{Introduction}

Monolayer graphene is a truly two-dimensional (2D) system and a zero-gap
semiconductor, where the electrons and holes behave as massless fermions with  unusual 
transport properties and display an anomalous quantum-Hall effect\cite{Nov05,Zha05}. 
In order to utilize their important electrical
features in certain applications, many efforts have been recently made to explore
the properties of lower-dimensional 
graphene nanostructures like nanoribbons\cite{Han07,Chen07,Sta08,Sol07,Wan08}.
By cutting graphene into a narrow ribbon structure, the electrons and holes 
are laterally confined to form a quasi-one-dimensional (1D) structure, similar to the case of
carbon nanotubes\cite{Wan08} and conventional semiconductor 
quantum wires\cite{Tho88,Ber86,Wha88,Wee88,Tho96,Tho98}. The formation of 1D 
subbands in graphene  structures can lead to the quantization
of conductance. One of the most interesting questions associated to this quantity is if the
conductance $G$ of the lower band deviates from that given by Landauer's formula\cite{Lan70} for 
non-interacting systems, $G=g_s\frac{e^2}{h}$ where the degeneracy factor is $g_s=4$ for graphene
and $g_s=2$ for conventional systems. 

Effects of the electron-electron interaction in conventional 1D systems were investigated by 
Tomonaga\cite{Tom50} many years ago, and since then a number of works have been published 
on this problem\cite{Lut63,Lu74,Fuk74,Luth74,Ape82,Kan92,Fur93,Oga94,Mal05}. One of the
remarkable predictions of these theories is that, even in the absence of scatterers 
such as impurities, the conductance G deviates from that given by Landauer's formula
as $G=\gamma g_s\frac{e^2}{h}$, where $\gamma$ depends on the interaction.
In an experiment on a GaAs-AlGaAs quantum wire, Reilly \emph{et al.} \cite{Rei01}
have found that $\gamma$ goes from 0.7 to 0.5 with increasing carrier density.
Another intresting effect provided by Tomonaga's model is the possibility for electron-electron interaction to
open a gap in the spectrum. This behavior has been recently experimentally verified in the contest of carbon nanotubes\cite{Desh09}.
          
Motivated by these experiments on quasi one-dimensional carbon-based systems,
in this work  we study the consequences of the electromagnetic interactions and quantum correlations between electrons and holes, 
inside  nanoribbons built from a single layer of two-dimensional graphene. For these systems, tight binding calculations of the 
1D subbands lead to a dispersion relation 
\be
E_n(k)=\pm\hbar v\sqrt{k^2 +(n+\alpha)^2\pi^2/w^2},
\ee
  $k$ is the longitudinal momentum, $v \simeq 10^6 m ~s^{-1} \simeq 1/300~ c$  is the band velocity and
$w$ is the nanoribbon width.
$n=0,\pm1,\pm2,....$ is
an integer for the subband index, and $0\le|\alpha|<0.5$ depends on the crystallographic orientation
of the GNR. The value of $\alpha=0$ holds in armchair nanoribbons (AGNR) families with N=3p+2,
where p is an integer and N is the number of dimer lines across the ribbon width.

In the following, we mainly focus on the computation of the conductance of graphene quantum wires in the case in which 
$n=0$ and $\alpha\simeq0$,  so that, in the absence of interactions,  the electron and hole $n=0$ bands form  gapless Dirac cones. 
To this end, we  develop a model motivated by  an effective  field theory (EFT)  based on an
expansion in $k/k_T$ (or $w/L$ where $L$ is the length of the system), where $k_T$ is the typical momentum of  electrons 
and holes propagating in the transverse direction. 

We show that the combined 
effects of the interaction
and of the low-dimensionality of the system drastically change the structure of 
the ground-state and of the spectrum of excitations,
which are typical of the two-dimensional graphene. In fact, the vacuum develops a condensate of electron-hole pairs and 
the spectrum of excitations turns out to be saturated by bosonic particle-hole collective modes.
As a consequence,  a gap opens up between valence and conduction band, turning the nanowire into a semiconductor: the AGNR behaves as a Mott insulator.
Interactions at distances of the order of few lattice spacing lead to a correction to Landauer's formula for  
the conductance, in the form $G=\gamma 4\frac{e^2}{h}$ with $\gamma=\frac{1}{1+d/\pi}$ and $-\pi<d<+\infty$. 

The paper is organized as follows. In section \ref{free}, we introduce the model  Hamiltonian  
for non-interacting  electron and holes and we calculate the free conductance of charge carriers.
In section \ref{local} we develop a model in which the screened Coulomb interaction is effectively local, and compute the corresponding conductance.  
In section \ref{reviewEFT} we use the  effective field theory formalism to construct the most general Lagrangian describing the interaction at distance of electrons and holes in the wire.  We show 
that, under suitable 
approximations, such a Lagrangian reduces to an exactly solvable model, whose solution is given in section \ref{solu}. 
The implications of these results on graphene physics will be presented 
in section \ref{spectrum}. Conclusions and perspective developments
are summarized in section \ref{conclusions}.

\section{Theory for GNR's without interaction}
\label{free}

The role of  dynamical electron-hole correlations in graphene quantum wires is expected to be very different than in mono-layer two-dimensional graphene. In fact, the free Dirac-like Hamiltonian for two-dimensional graphene has been shown to be a fixed point of the Renormalization Group~(RG) of the corresponding EFT~\cite{RG2D}. This implies that perturbative coupling constants of such an EFT can be made arbitrarily small by applying RG transformation. 
Hence, the free Dirac Hamiltonian can be used as a good starting approximation for discussing the dynamics of electrons and holes in a graphene layer, with interactions providing at most, small logarithmic corrections. On the other hand,
in \cite{PRB1}  we have shown that the lowest-order interaction 
terms  the EFT describing quasi one-dimensional graphene have dimensionless coupling  constants, and therefore appear at  the same order in the EFT expansion of the free kinetic terms. 
This implies that interactions can never be neglected, and in general are expected to shape the physics of such quantum wires. 

In view of such  considerations it is instructive to first consider as a bench mark  the quantum motion near the K point 
of the electrons in the conductance band and holes in the valence band of GNR's in the limit in which all interactions are neglected. Such an analysis allows to clarify which properties of the wire are induced by the dynamical correlations. 

The second-quantized Hamiltonian leading to the desired free single-particle spectrum
of the lowest energy band 
\be
\omega(k)=  \pm \hbar v \sqrt{ k^2 + {M^2v^2\over \hbar^2}} 
\label{disp}
\ee
is of course
\be
\label{H0}
H_0 = v \hbar \int d x ~ \psi^\dagger (t,x)\left( -i \sigma_1 \partial_x +{M v\over\hbar}\right) \psi(t,x),
\label{Ham}
\ee
where $\psi(t,x)$ is the fermion field operator. In the following, 
we shall work in a natural system of units for this problem, in which 
 $\hbar=v=1$. Note that, in such units, the speed of light is  $c=1/\beta\simeq 300$.

In order to exploit the formal analogy with the relativistic Dirac theory, 
it is convenient to introduce  position contravariant  vectors 
\be
\tilde x^\mu=(v t, x) = ( t, x),
\ee  
momentum contravariant vectors  
\be
\tilde p^{\mu}=(E/v, p)=(E,p),
\ee  
and the metric tensor as $g_{\mu \nu}=\textrm{diag}[1,-1]$. 
In addition, let $\gamma^\mu$ and $\gamma^S$ be  $2\times 2$ matrices obeying the usual Dirac algebra:  
\be
\{\gamma^\mu, \gamma^\nu\} &=& 2  g^{\mu \nu}\\
  \{\gamma^S, \gamma^\mu\} &=& 0
 \ee
and, $(\gamma^S)^2 =  1$.
For example, one may choose a representation in which
\be
\gamma^0 &=&  \sigma_3\nonumber\\
\gamma^1 &=& i \sigma_2 \nonumber\\
\gamma^S &=&  \gamma^0 \gamma^1= \sigma_1 
\label{gamma}
\ee
Note that with this choice  one has $\gamma^\mu \gamma^S =  - \epsilon^{\mu \nu} \gamma_\nu.$

Using such a set of definitions, the action associated to the free "Dirac" Hamiltonian (\ref{Ham}) 
can be cast in the familiar form
\be
S_0 =   \int d^2 \tilde x ~  \bar{\psi} ~\left(i ~ \tilde {\ds\partial}-M\right) ~ \psi.
\ee

Furthermore, in the presence of a classical external field $A^\textrm{ext}_\mu$,
the action of free electrons and holes inside the system is given by
\be
\label{Ssh}
S= \int d^2 \tilde x ~\bar \psi\, \left(i \tilde{\ds D}-M\right) \,\psi ,
\ee
where the covariant derivative is defined as
\be
\label{covD}
\tilde D_\mu=\tilde{\partial}_\mu + i e A^\textrm{ext}_{\mu}.
\ee

The electric conductivity of the GNR  described by the action (\ref{Ssh}) is given by
\be
\Re\left[\sigma(\omega, q)\right] = \frac{1}{E^{ext}(\omega, q)} ~\Re \int d^2 
\tilde x e^{i (\tilde \omega \tilde{x}_0- \tilde q \tilde x)} \langle j_1(\tilde x) \rangle_{ A^{\mu}},\nonumber\\
\ee
where $ A^{\textrm{ext}}_{\mu}(\tilde x)=(\Phi(\tilde x),0)$  is
the potential of a weak external electric field, $E^\textrm{ext}$.

Applying the linear response theory, one immediately finds
\begin{equation}
\label{inizio}
\int d^2 \tilde x~ e^{i \tilde q\cdot \tilde x} ~\langle j^1(\tilde x)\rangle_{ A^{ext}_\mu} 
=  i~(i \Pi^{1 0}(\tilde q))~  A_0^{ext}(\tilde q),
\end{equation}
where $\Pi^{\mu \nu}(\tilde q)$ is the vacuum polarization tensor, defined as:
\be
i \Pi^{\mu \nu}(\tilde q) = \int d^2 \tilde x ~e^{i \tilde q \cdot \tilde x} 
~\langle \Omega| T[ j^\mu(\tilde x) j^\nu(0)] |\Omega \rangle.  
\ee
In the non interacting  model, this matrix element can be computed exactly and reads
\be
\label{pol}
i \Pi^{\mu \nu}( \tilde q) =- \frac{e^2}{\pi} 
\left(g^{\mu \nu}- \frac{\tilde q^\mu \tilde q^\nu}{\tilde q^2}\right)I(q)~,
\ee
where 
\be
\label{int}
I(q)=-q^2\int_0^1 d~x\frac{x(1-x)}{M^2-x(1-x)q^2}~.
\ee

This result can be used to readily obtain the conductivity of the system. 
After restoring the appropriate powers of $v$ and $\hbar$, we find our final results:
\be
\Re\left[{\sigma(\omega,  q)\over L}\right] &=& \Im 
{e^2\over\pi\hbar}{\omega/v\over(\omega/v+ q)(\omega/v- q)}I(q)
\nonumber\\
&=&\frac{e^2}{2 \hbar}\delta(\omega/v -q)I(\omega/v),
\label{cond}
\ee

and by taking the Fourier transform\cite{Kaw95}  
\be\label{fine}
\Re\left[{\sigma(\omega, x)\over L}\right]=\frac{e^2}{2\pi \hbar}\cos \left({\omega x\over 
v}\right)I(\omega/v)~.
\ee
When the mass term $M$ is zero and the electron and hole bands are gapless Dirac cones,
as happens for armchair GNR's with N=3p+2, 
in the limit $\omega\to0$, we get for the DC conductance $G$  
(defined as the $\omega\to0$ 
limit of 
$\Re\left[{\sigma(\omega, x)\over L}\right]$) at zero chemical potential:
\be
G={e^2\over h}~.
\label{cond2}
\ee
This result must then be multiplied for the factor $g_s=4$, to account for the spin and sub-lattice degeneracy:
\be
G\rightarrow G= g_s~{e^2\over h}.
\ee

When there is a band gap, i.e. $M\ne0$, 
we expect that the same result holds when $M$ is small with respect to the Coulomb energy  
necessary to induce transport in the system.
 Otherwise, the DC conductance at zero chemical potential is zero.  

This discussion can be straightforwardly generalized to the case of finite electron chemical potential. 
In  general, in the formalism of quantum field theory, the transition to finite density is achieved by adding a term to the zero-density action in the form 
\be
\label{chemicaldiff}
S[\psi, \bar \psi] \rightarrow S[\psi, \bar \psi]  - i \mu \int d^2 \tilde{x} \psi(\tilde{x})\gamma_0\psi(\tilde{x})
 \ee
Hence, in the free-theory case, the finite-density effect is simply that of shifting the $\tilde p_0$ component  which enters in the free fermion propagator by the electron's chemical potential $\mu$: $\tilde{p}_0 \to \tilde{p}_0+\mu$.   
As a consequence, the first branch-cut singularity of $I(\tilde{q}^2)$ in the  complex plane is translated to the point $\tilde q_0=M-\mu$, on the real axis. This means that, even for arbitrarily small Coulomb energies, the regime of finite conductance is reached as soon as the electron density $\mu$ becomes equal or larger than the gap $M$, as expected.  

In the next section we will take into account the effect of the interaction between electrons and holes on 
$G$ in the case of $M=0$.

\section{A model with local current-current interactions}
\label{local}

Let us now begin our study of the effects of the  interactions between electrons and the holes. In this section, we consider a model which emphasizes the consequence of the screening of the Coulomb force inside the wire, and completely neglects the interaction at distance. 

\begin{figure}[t!]
\includegraphics[width=8cm]{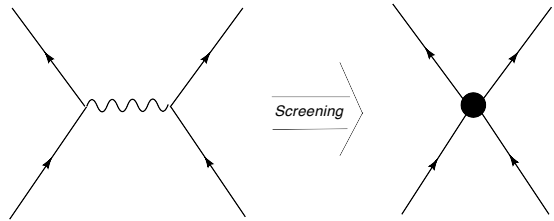}
\caption{If the typical momenta exchanged by fermions in the wire is much smaller than the screening length of the electro-static interaction, the interaction can be replaced by an effective local vector-vector vertex. }
\label{screen}
\end{figure}

If the electro-static interaction between electrons and holes is short-ranged, then the very low-energy electro-dynamics 
can be described by an effective  vector-vector interaction---see Fig. ~\ref{screen}--- 
\be
\label{th}
\mathcal{L}_{int} = \frac{d}{2}~ (\bar \Psi \gamma_\mu \Psi)~(\bar \Psi \gamma^\mu \Psi),
\ee
where the coupling constant $d$ is inversely proportional to the inverse of the screening mass $m_s$, i.e.
\be
\frac{d}{2}=e^2/m_s^2.
\ee
Notice the analogy with the Fermi theory for weak decays, in which the vector-boson mediated weak-interaction is replaced  by an effective local axial-vector coupling. 

In our 1+1 dimensional effective theory, the  choice of interaction (\ref{th}) defines the massless Thirring Model, i.e.
 \be
 \label{fermionthirring}
 S_{Th}= \int d^2 \tilde x ~\bar \psi\, i \tilde{\ds \partial} \,\psi - \frac{d}{2} \left (\bar \psi \gamma_\mu \psi\right)^2.
 \ee
  The calculation of the conductance of the wire in such a model can be carried out analytically, even in the non-perturbative regime. 

As shown in\cite{Coleman, Abd, zinn, Faber} it is possible to explicitly reformulate the massless Thirring Model in terms of a massless bosonic free theory.  In the present context,  the bosons of such a theory can be physically interpreted as a electron-hole bound states. 
The fact that the spectrum of the wire in the Thirring Model is  gapless implies that short-ranged vector correlations cannot turn the metallic armchair nanoribbon into a semiconductor.

Let us now compute the conductance of the wire. Following\cite{zinn} we bosonize the system and map the (\ref{fermionthirring}) into
\be
S=\frac{1}{2}\int d^2\tilde x \left[(\tilde\partial_\mu\theta)^2+\frac{1}{d}(\tilde \partial_\mu\chi)^2-\left(\frac{1}{d}+\frac{1}{\pi}\right)(\tilde \partial_\mu\phi)^2\right],\nonumber\\
\ee
where $\theta$ is the scalar field obtained through the bosonization of the fermionic fields, while $\chi$ and $\phi$ are the boson fields introduced using the
auxiliary vector field technique\cite{faber2}.

The bosonization dictionary for the massless Thirring model\cite{Coleman, zinn} states that the vector current is represented as follows
\be
J_\mu=\bar{\Psi}\gamma_\mu\Psi \to \frac{1}{2\pi}\epsilon_{\mu\nu}\partial_\nu\theta.
\ee
Hence,  it is straightforward to calculate the response of the system to an external  field, in linear response theory. The current-current correlation function reads
\be
\langle J_\mu(p)J_\nu(-p)\rangle=\left(g_{\mu\nu}-\frac{p_\mu p_\nu}{p^2}\right)\frac{e^2}{\pi(1+d/\pi)}.
\ee

Using the procedure (\ref{inizio})-(\ref{fine}) we can compute the quantum conductance for the AGNR. The result is
\be
G= g_s~\frac{e^2}{h(1+d/\pi)}.
\ee
Notice that this results  differs from Landauer's formula by the factor\footnote{This result holds in the chirally conserved phase, i.e. $d>-\pi$. We expect that a modification of Landauer's formula
will exists even in the chirally broken phase~\cite{Faber}, but this scenario will not be discuss here.}
$1/(1+d/\pi)$ which is due to the presence of  short-ranged correlations.

\section{Including the Effects of the Interactions at Distance}
\label{reviewEFT}

In the previous section, we have developed a phenomenological model in which the Coulomb interaction inside the wire was assumed to be effectively zero-ranged. Such a model becomes reliable only if the typical momenta of the electrons and holes propagating in the wire are much smaller than the inverse screening length. 

We now want to improve on such an approach, by defining an alternative model in which the range of the interaction is kept finite. 
To this end, we rely on  the effective field theory (EFT) formalism, which represents a powerful tool to describe the low-energy dynamics of arbitrary closed systems ---for a  pedagogic introduction to EFT's see \cite{howto}, for a more technical treatment see e.g. \cite{Manho}---. 

Any time a physical system is characterized by a large separation in its relevant energy-momentum scales, the low-energy observables are expected to be insensitive to the details of the physics which involves only the hard degrees of freedom, with energy and momenta above the gap.
In this case, the EFT formalism can be used  to systematically construct a model Lagrangian which describes the dynamics of the relevant low-energy degrees of freedom.  
In general,  the EFT  is  much simpler to solve than the corresponding (more) fundamental theory, as it is formulated in terms of fewer degrees of freedom.  
In addition, the advantage of the EFT description is that its  Lagrangian can be systematically constructed, starting from  symmetry arguments. 
Indeed,  Weinberg theorem\cite{Weinberg} implies that one should consider the most general Lagrangian, compatible
 with the symmetry properties of the underlying fundamental theory and with the fundamental principles of quantum field theory:
\be
\label{EFT}
\mathcal{L} = \sum_{i=1}^{\infty} c_i(\Lambda) ~\hat{O}_i.
\ee
In this Eq., $\hat O_i$ are field dependent local operators, $\Lambda$ is an ultra-violet cut-off, chosen in the gap between the high-energy and the low-energy modes and $c_i(\Lambda)$ are the (running) coupling constants. 
The price to pay in the EFT approach is that the Lagrangian contains in principle an infinite number of new unknown parameters $c_i(\Lambda)$, which implicitly embody the information about the ultra-violet physics, above the cut-off. Such coefficients have to be calculated from the underlying microscopic theory, or have to be fitted from experiment.  

Clearly, the Lagrangian (\ref{EFT}) does not yet represent a physical theory, since it depends on an infinite number of parameters. 
On the other hand, the presence of a large separation in the energy-momentum scales assures that, in order to compute observables to any finite accuracy, one needs to specify only a  finite number of such effective parameters\footnote{The situation is completely analog to the multi-pole expansion of classical electrodynamics: the electro-magnetic  field at distances much larger than the size of the source can be determined with arbitrary accuracy from a finite number of multipole coefficients.}. Typically, this corresponds to the coupling of the lowest-dimensional operators $\hat O_i$. In the case of  perturbative theories, one can show that higher dimensional interaction operators lead in general to higher order contributions\footnote{At most,  perturbative quantum fluctuations can lead to corrections to the naive dimensions of the fields.} in the expansion of the observables in power of $p/\Lambda$. 
 Hence, any desired accuracy can be reached by considering effective Lagrangians with only a finite number of effective interactions. 

Let us now apply the EFT formalism to our specific case, in which the natural cut-off scale is provided by the transverse momentum in the wire, 
$\Lambda=k_T$.  
We are interested in an effective field theory describing only the dynamics of low-energy degrees of freedom inside the wire, i.e. 
the electrons and holes propagating along the longitudinal direction. 
Hence, in our approach, we propose  a purely $1+1$ dimensional description of the electronic properties of the wire, using the EFT technology to build an effective  Lagrangian which simultaneously takes into account for the short- and long-ranged correlations induced by
 the presence of the  transversal confinement and of the interactions between electrons inside the wire.

The only way to include interactions at distance, while respecting the symmetry, causality, unitarity and \emph{local} charge conservation constraints is to introduce
an effective gauge-boson field $a_\mu$, which we shall call the pseudo-photon, owing to the formal analogy with QED.
It is important to stress that this is an \emph{effective} field, which does not represent physical photons propagating inside the wire. Its role is to mediate the long distance interactions arising from the interplay of electro-magnetic coupling and transverse confinement,  in a way that charge conservation is enforced locally, in the wire.   
In addition, since we are interested in the response to an external electric field, we include into our effective Lagrangian the coupling with a (\emph{physical}) external  electro-magnetic field $A^\textrm{ext}_\mu$.
On the other hand, we neglect the dynamics of photon radiation, from the electrons in wire, as this a relativistic effect and arguably very small, in the limit of low electron density.  

The pseudo-photon field $a_\mu$ propagates at a speed $v'=1/\beta'\le 1/\beta$, which does not in general correspond to that of the electrons and holes.
 This is equivalent to say that we can define a gauge in which the free classical Euler-Lagrangian Eq.s for $ a_\mu$ are:
\be
\label{freewave}
\left(\beta'^2 \frac{\partial^2}{\partial t^2} -\frac{\partial^2}{\partial x^2} \right)~ a_\mu = 0.
\ee
This can be achieved by considering a free action in the form
\be
\label{freepphoton}
S_0[\tilde a_\mu] = \int d^2\tilde x~\left(-\frac{1}{4}~\tilde f_{\mu \nu} \tilde f^{\mu \nu}\right),
\ee 
where 
\be
\tilde f_{\mu \nu} = \tilde{\partial}_\mu \tilde a_\nu - \tilde{\partial}_\nu \tilde a_\mu,
\ee
and 
\be
\tilde{a}_0 &=& a_0,\\
\tilde{a}_1 &=& \beta^\prime ~a_1.
\ee
The action (\ref{freepphoton}) leads to the free wave Eq. (\ref{freewave}) upon imposing the modified Lorentz gauge condition,
\be
\tilde \partial_\mu \tilde a^\mu =0. 
\ee 
Clearly, since the pseudo-photon and the massless fermions propagate at different velocities, the resulting  theory will not be invariant under any "Lorentz"-like symmetry. 

The starting point to construct an effective Lagrangian which conserves locally the chage of the fermions is to introduce a gauge invariant coupling of the pseudo-photons and fermions:
\be
\tilde D_\mu \equiv \tilde{\partial}_\mu + i g~ a_\mu + i e A^\textrm{ext}_{\mu}.
\ee

Let us now perform a naive dimensional analysis to identify the lowest order operators in our EFT. 
In $1+1$ dimensions, the fermion field has mass dimension $1/2$, while the $a_\mu$ 
and $A_\mu$ field have mass dimension $1$.
The  lowest-dimensional gauge invariant and "Lorentz"-invariant operators are therefore
\be
&&\bar \psi\, i \tilde{\ds D} \psi, ~m(\Lambda) \bar \psi \psi,~ (\bar \psi O \psi)^2,   \qquad( O=1, i \gamma_s, \tilde \gamma_\mu,\tilde \gamma_\mu \gamma_S) \nonumber\\
&&  \bar \psi  \sigma_{\mu \nu} \tilde f^{\mu \nu}  \psi,   \bar \psi  \sigma_{\mu \nu}  F^{\mu \nu}  \psi  \nonumber\\
&& \tilde f_{\mu \nu} \tilde f^{\mu \nu},~  \bar \psi\, i \tilde \partial_\mu \tilde f^{\mu \nu} \gamma_\nu \psi, \bar \psi\, i \tilde \partial_\mu  F^{\mu \nu} \gamma_\nu \psi.
\label{list}
\ee
The first, second and third lines contain operators of dimension 2, 3 and 4, respectively. 

Form this point ahead we will consider a special configuration for the AGNR, which geometrically assures the disappearance of the mass term, as in planar graphene, due
to the contact between valence and conduction bands. As shown in\cite{nakada}
this hypotheses is fulfilled in 3p+2 family of AGNR, that represents $1/3$ of the whole possible armchair nanoribbons.


The role of the contact terms $(\bar \psi O \psi)^2$ is to mimic the ultra-violet
physics which sets in when electrons and holes interact at a distance of the order of the inverse cut-off, i.e. of the transverse size of the wire, 
$\simeq  1/k_T$. 
The interaction between electrons and holes separated by distances much larger than the inverse cut-off is mediated by the coupling with the pseudo-photon field,  e.g. the $\bar \psi\, i \tilde{\ds D} \psi$ term.

The exact non-perturbative solution of the theory described by the EFT Lagrangian containing all the terms in (\ref{list}) is of course a formidable problem. 
On the other hand, our main purpose is to study the combined effect of long-range and short-range correlations on the electronic properties
of the wire. From such a stand point,  we choose to retain in our model Lagrangian only the  vector-vector contact interaction term $(\bar \psi \gamma_\mu \psi)^2$, the minimal coupling term $\bar \psi\, i \tilde{\ds D} \psi$ and the pseudo-photon kinetic term $\tilde f_{\mu \nu} \tilde f^{\mu \nu}$. 

In conclusion, the effective action of our model for the internal quantum electrodynamics of the wire reads
\be
\label{S}
S \equiv  \int d^2\tilde x ~\Big[\bar \psi\, i \tilde{\ds D} \,\psi - \frac{d}{2}\big(\bar{\Psi}\gamma_\mu\Psi\big)^2 -  \frac{1}{4} \tilde f_{\mu \nu} \tilde f^{\mu \nu}~\Big].
\ee
We emphasize the fact that such an action was obtained by retaining only some of the effective couplings. This choice unavoidably introduces some model dependence in our calculations. On the other hand, the gain is that the model defined by (\ref{S}) is exactly and analytically solvable, even in the non-perturbative regime.  In the following, we shall refer to the model define by the action (\ref{S}) as to the Schwinger-Thirring model.

\section{Exact Non-Perturbative Solution of the Schwinger-Thirring Model}
\label{solu}

In this section, we discuss the exact analytic solution\footnote{ In order to keep the notation as simple as possible, all the derivations reported in this section correspond to the choice $\beta'=1.$ The generalization to arbitrary values of $\beta'$ is straightforward, although the notation is considerably more involved.} of the Schwinger-Thirring model ---see also the discussion in \cite{cinese}---.  

\subsection{Solution of the Schwinger Model}
\label{Schwingsolu}

Let us begin by reviewing the solution of the pure Schwinger model, i.e. for $d=0$, which can be found in standard quantum field theory textbooks, such as \cite{zinn}. 
A remarkable feature of such a theory is that its  classical action is invariant under \emph{two} independent  transformations, defined by  the following different  types of local rotations of the fermion fields
\be
\label{rotgauge}
\psi(x)&\to& e^{i \chi(x)} \psi(x)  \qquad  \bar \psi(x)\to \bar \psi(x) e^{-i \chi(x)}\\
\label{rotchiral}
\psi(x)&\to& e^{i \gamma^S \Phi(x)} \psi(x)  \quad  \bar \psi(x)\to \bar \psi(x) e^{i \gamma^S \Phi(x)}
\ee
and by the corresponding gauge transformations:
\be
\label{gauge}
{a}_\mu &\to&{a}_\mu - \frac{1}{g}{\tilde\partial}_\mu \chi,\\
\label{chiral}
{a}_\mu &\to&{a}_\mu + i\frac{1}{g}~\epsilon_{\mu \nu}{\tilde\partial}^\nu \Phi.
\ee

The dynamical consequences of such a symmetry become evident once one parametrizes the photon field degrees of freedom as
\be
a_\mu = \frac{1}{g} \left({\tilde\partial}_\mu \chi - i\epsilon_{\mu \nu}{\tilde\partial}^\nu \phi~\right)
\label{subs}
\ee
and  re-expresses the path integral in terms of the fermion fields $\psi, \bar \psi$ and of the $\chi$ and $\phi$ fields.

The invariance of the classical action defines a classical symmetry of the system. 
Such a symmetry is realized also at the quantum level only if the full path integral remains unchanged under the transformations (\ref{rotgauge})-(\ref{rotchiral}), and (\ref{gauge})-(\ref{chiral}).
To verify if such a  condition is realized, let us analyze the transformation properties of the fermionic measure 
$\mathcal{D}\psi \mathcal{D} \bar \psi$. It is  possible to show that the gauge symmetry (\ref{rotgauge}) leaves invariant the fermionic measure and therefore the symmetry defined by the transformation (\ref{rotgauge})-(\ref{gauge}) is respected also at the quantum level. 
On the other hand, under the chiral gauge transformation (\ref{rotchiral})-(\ref{chiral}) one has 
\be
\mathcal{D} \psi \mathcal{D} \bar \psi \to \mathcal{D} \psi \mathcal{D} \bar \psi \mathcal{J}^{-2}, 
\ee
where 
\be
\mathcal{J}^{-2}= \exp\left[-2 \int d^2 \tilde x~\frac{ 1}{g^2} ~ {\tilde\partial}^\mu \phi~  m^2{\tilde\partial}_\mu~\phi\right]
\ee
is the functional Jacobian determinant of the chiral transformation (\ref{rotchiral}) and $m=g/\sqrt{\pi}$ is the so-called Schwinger mass.
Hence, the chiral gauge symmetry is said to be "anomalous", i.e.  broken at the quantum level. 

As a result of the chiral anomaly, the path integral of the Schwinger  model can be written as
\be
Z= \int \mathcal{D}\psi  \mathcal{D}\bar \psi \mathcal{D} \phi ~\exp\left[ i S_\textrm{shw}[\psi, \bar \psi, \phi]\right],
\ee
where the Schwinger action $S_\textrm{shw}$ is defined as
\be
\label{Shw}
&& S_\textrm{shw}[\psi, \bar \psi, \phi]  =
\int d^2 \tilde x  ~\left[\bar \psi\, i\hf{ \tilde{\partial}} \,\psi  \right. \nonumber\\
 &-&\left.\frac 1 2 ~\frac{ 1}{g^2} ~ {\tilde\partial}^\mu \phi~ \Big( \tilde\partial^2- m^2 \Big)~{\tilde\partial}_\mu~\phi\right].
\ee

As shown in \cite{zinn},
this partition function can be mapped into a purely bosonic one, which reads
\be
S_\textrm{shw}[\theta, \phi] = \frac{1}{2} \int d^2 \tilde x \left[\left(\tilde\partial_\mu\theta\right)^2-m^2\theta^2\right].
\ee
Notice that this action is quadratic in the boson fields, so every correlation function can be computed exactly.

\subsection{Solution of the Schwinger-Thirring Model}

We are now in a condition to discuss the solution  of our Schwinger-Thirring model~ (\ref{S}).
Applying the chiral and scalar gauge transformations (\ref{rotgauge})-(\ref{gauge}) and  (\ref{rotchiral})-(\ref{chiral}), and evaluating the corresponding anomalous shift in the fermionic measure, the action transforms into
\be
\label{St2}
 S_\textrm{shw}[\psi, \bar \psi, \phi]  =
&& \int d^2 \tilde x ~\left[\bar \psi\, i\hf{ \tilde{\partial}} \,\psi - \frac{d}{2} \left (\bar \psi \gamma_\mu \psi\right)^2 +
\right. \nonumber\\
 &-&\left. \frac{1}{2} ~\frac{1}{g^2} ~ {\tilde\partial}^\mu \phi~ \left( \tilde\partial^2- m^2 \right)~{\tilde\partial}_\mu~\phi~\right].~\mbox{}\qquad
\ee
As in the case of the Schwinger model, the strategy to solve this theory is to define a bosonization scheme such that the resulting formulation of the path integral is purely Gaussian. 
Such a procedure involves  determining the bosonized representation of the vector current, $\bar \psi \gamma_\mu \psi$. The procedure is illustrated in the appendix~(\ref{appST}).
The result of such an analysis is that  under bosonization  the vector current operator becomes  
\be
\bar{\Psi}\gamma_\mu\Psi\rightarrow-i\frac{1}{\sqrt{\pi}}\epsilon_{\mu\nu}\tilde\partial_\nu\theta.
\ee

Having established the  "bosonization dictionary"\cite{Adam, Abd} for the fermionic current, it is immediate to obtain the bosonized version of the Schwinger-Thirring model. Indeed, Eq. (\ref{St2}) transforms into
\be
\label{bosonST}
S_\textrm{st}[\theta, \phi] = \frac{1}{2} \int d^2 \tilde x\left[ \left(1+\frac{d}{\pi}\right)\left(\tilde\partial_\mu\theta\right)^2 - m^2\theta^2\right],
\ee
with $m^2=g^2/\pi$.
Formally, this action has the same form of the original Schwinger model bosonic action. The only difference is in the  factor $(1+d/\pi)$, which rescales the kinetic term. This factor contributes to
correlations functions and provides corrections to the mass of the Schwinger boson, but does not modify the qualitative structure of the ground-state and of the excitation spectrum.

\section{Properties of the GNR in the Schwinger-Thirring Model} 
\label{spectrum}

In section \ref{reviewEFT}  we introduced the  Schwinger-Thirring model for  the electrodynamics of the graphene wire. In the previous  section, we have  shown that such a model can be mapped into a modified Schwinger model. The effect of the short-range vector-vector interaction which are present in our model is  absorbed into the coefficient  $(1+d/\pi)$, which multiplies the kinetic energy term. 
Hence, using the fact that the Schwinger model is exactly solvable, we are finally in a condition to predict some important properties for the wire.

\subsection{Ground-state structure and bosonization of the spectrum of excitations}

As in the original Schwinger model, the vacuum  is characterized by a finite fermion condensate
\be
\langle\Omega|\bar{\Psi}\Psi|\Omega\rangle\ne 0.
\ee
In the context of graphene nanowire physics, this result implies that, even in the presence of short-distance vector-vector interaction, the electron-hole pair density in the vacuum is not zero.

This effect is due to the anomalous breaking of the chiral symmetry, which is also responsible for the bosonization of the spectrum. The
Schwinger-Thirring model contains no fermion excitations, but only an arbitrary number of free fermions-antifermions bound states, with mass $M=g^2/\pi(1+d/\pi)$. 
In the context of graphene physics,  this means that the spectrum of excitations of the nanowire starts with a {\it single collective electrons-holes mode} with dispersion relation
\be
\omega(k)=\pm\hbar v \sqrt{k^2+\frac{4 g^2 \alpha/\beta}{e^2 (1+d/\pi)}}
\ee
and contains a continuos of multi-boson excitations, starting at the two-boson threshold.
Additional thresholds for multi-boson excitations are located at $n m$, with $n=3,4,5..$.
In these formulas, we have restored the constants $v$, $c$, $\hbar$ and $\alpha={e^2\over4\pi\hbar c}\simeq1/137$ is the fine structure  constant.

\subsection{Conductance of the graphene wire}

The calculation of the conductance of the  wire in linear response theory in the presence of electron-hole interactions can be performed following the same steps taken in the free  case. 
It is important to recall that  the model we are considering contains two distinct vector fields: the effective pseudo-photon field  $a_\mu$ and the physical external photon field $A_\mu$. The conductance is stimulated by the latter field, while the long-range dynamical correlations are generated by the former field. 
Consequently, the coupling of such two fields with the fermion field $\psi$ is parametrized by different coupling constants, $g$ and $e$. 
Note that this is not the case in the original Schwinger model, in which one considers only one type of coupling of the fermions to the vector field ---see e.g. \cite{Peskin}---. 
As a consequence, the current-current correlation function which enters in the definition of the conductance in our model is not the same correlation function which enters in the Dyson series associated to the pseudo-photon mass renormalization\footnote{Note that, in our previous work \cite{PRB1} this point was overlooked. The calculation of conductance reported in such a work  but holds only for values of the chemical potential larger than the Schwinger mass (i.e. of the dynamically generated semi-conductor's gap).}.

The current-current correlation function  $\Pi^{\textrm{sh}}_{\mu\nu}$ of the Schwinger system to an external vectorial perturbation was  calculated by Schwinger in his original work\cite{schwref}, but  the bosonisation technique offers
a different way to obtain the same result. By definition we have
\be
&&i \Pi^{\textrm{sh}}_{\mu\nu}(\tilde x,\tilde y)= \frac{\delta}{i \delta A_\mu(\tilde x) ~i\delta A_\nu(\tilde y)}\Bigl|_{A_\mu=0}~ \log Z[A_\mu],\\
&&Z^{\textrm{sh}}[A_\mu]= \int \mathcal{D}  a_\mu \mathcal{D}\bar{\Psi}\mathcal{D}\Psi~e^{iS_{sch}+ i~e~\int d^2\tilde x \bar{\Psi}(\tilde x)\gamma_\mu\Psi(\tilde x)A^\mu(\tilde x)}\nonumber\\
\ee
applying the bosonization technique, we find 
\be
i\Pi^{\textrm{sh}}_{\mu\nu}(\tilde x,\tilde y)=-\frac{e^2}{\pi}  \la \epsilon_{\mu\sigma}\tilde \partial_\sigma \theta(\tilde x) \epsilon_{\nu\tau} \tilde\partial_\tau \theta( \tilde y)\rangle.
\ee

After analytically continuing back to real time, the Fourier transform of such a result  is
\be \label{errata}
i \Pi^{\textrm{sh}}_{\mu\nu}(\tilde q)=-\frac{e^2}{\pi}\left( \tilde q^2g_{\mu\nu}-\tilde q_\mu \tilde q_\nu\right)\frac{1}{\tilde q^2-m^2},
\ee
where $m$ is the Schwinger mass.  The effect of long-range correlations on current-current correlations can be read-off by comparing this formula its free counterpart, Eq. (\ref{pol}).

By following exactly the same procedure one can compute the current-current correlation function in the complete Schwinger-Thirring model, i.e. using the action (\ref{bosonST}).  We find 
\be
i\Pi^{\textrm{st}}_{\mu\nu}( \tilde q)=
-\frac{e^2}{\pi}\frac{1}{1+d/\pi}\left( \tilde q^2 g_{\mu\nu}- \tilde q_\mu  \tilde q_\nu\right)\frac{1}{ \tilde q^2-m^2}.
\ee
We note that the existence of a single pole implies that the conductance is entirely saturated by the propagation of confined electron-hole bound states (Schwinger bosons), and not by (quasi) free electrons and holes.  

Since the system develops a gap, the conductance becomes different from zero for electron densities larger than the gap, i.e. for $\mu\ge m$.
If is this the case, the procedure used from Eq. (\ref{pol})  to Eq. (\ref{cond2}) can be repeated to find the quantum of conductance. The result is 
\be
G=g_s \frac{e^2}{h(1+d/\pi)},
\ee
where we have included the degeneracy factor. 

The present analysis has illustrated how the transport properties of the wire are determined by  two different mechanisms  which are related to the long- and short- range part of the interaction, respectively. Interactions  at distance  induce a gap in the spectrum,  implying the insulating property of the nano-wire. However, long-ranged interactions alone
are not able to modify the value of  conductance's quantum. Corrections to the free-theory value $4 e^2/h$  value are completely due to short-ranged interactions.

\section{Conclusions}
\label{conclusions}
In this work, we have studied the effect of the interactions between electron and holes on the conductance $G$ 
of quasi-one-dimensional graphene systems at zero temperature. We first considered the case in which all the
 interactions 
are absent and there is a gap in the electron and hole bands, described by an effective mass term $M$ 
in the free Hamiltonian. In this case, {when the Coulomb energy of the field applied to the ends of the wire is larger than the gap $M$, we recover the well known result $G=4 e^2/h$.  On the other hand, when such a Coulomb energy is much smaller than the gap, we obtain a finite conductance $G=4 e^2/h$ only  
when the chemical potential  $\mu$  becomes equal or larger than $M$, as expected. }

We have then taken into account the effect of the 
interaction by developing a model inspired by an effective field theory based on an expansion in $k/k_T$, where $k$ and $k_T$
are the momenta in the longitudinal and transverse directions, respectively. We have shown that long-range interactions dynamically generate a gap in the spectrum, turning the GNR into a Mott insulator. In addition,  short-range interactions lead to a renormalisation of the free theory
result for the conductance.
Once both types of interactions are taken into account, one obtains that when the Fermi energy exceeds  the dynamically generated gap, the conductance is  $G=4 e^2/h(1+d/\pi)$, where $d$ is the short-range interaction strength.



A possible development of the present work would be to investigate 
how the properties of the wire change as a function of the temperature.
Also, our field-theoretic approach can be easily implemented in the case of Carbon Nanotubes. 

\appendix
\section{Bosonization of the Schwinger-Thirring Lagrangian}
\label{appST}

Let us consider the theory defined by the partition function
\be
Z_{\bar{\Psi},\Psi,a_\mu}[A_\mu] &=& \int \mathcal{D}\Psi \mathcal{D}\bar{\Psi}\mathcal{D} a_\mu~\exp\Big[-\int  d^2 x~\frac{1}{4}f^{\mu\nu}f_{\mu\nu}+\nonumber\\
&-& \bar{\Psi}
\left(\ds\partial+ie\ds a+ie\ds A\right)\Psi
\Big]. 
\ee
Notice that, for sake of simplicity and without loss of generality, we have dropped the "tilde" notation and we have assumed the same coupling constant $e$ for both the dynamical $a_\mu$ and the external $A_\mu$ fields. 
  
To solve this theory, we can proceed in the same way as section (\ref{Schwingsolu}), applying the gauge transformation also 
to the external field $A_\mu$. Hence, we define: 
\be
a_\mu=\frac{1}{e}(\partial_\mu\chi-i\epsilon_{\mu\nu}\partial_\mu\phi),
\ee
\be
A_\mu=\frac{1}{e}(\partial_\mu\alpha-i\epsilon_{\mu\nu}\partial_\nu\beta).
\ee
The corresponding path integral becomes
\be     
&&Z_{\bar{\Psi},\Psi,a_\mu}[\alpha, \beta] = \int \mathcal{D} \Psi \mathcal{D}\bar{\Psi}\mathcal{D}\chi \mathcal{D}\phi \nonumber\\
&& \exp\Bigr\{-\int d^2 x \frac{1}{2e^2}\partial_\mu\phi\partial^2\partial^\mu\phi \nonumber\\
&-& \bar{\Psi}\Bigr[\ds\partial+
\gamma^\mu\Big(-\partial_\mu\left(\chi+\alpha\right)-i\epsilon_{\mu\nu}\partial_\nu\left(\phi+\beta\right)\Big)\Bigr]\Psi\Bigr\}.\nonumber\\
\ee
In the presence of the external field, the application of the gauge and chiral gauge rotations leads to two anomalous terms, which are 
related to the $\phi$ and $\beta$ fields, respectively:
\be
-\frac{1}{2\pi}\left(\partial_\mu\phi\right)^2-\frac{1}{2\pi}\left(\partial_\mu\beta\right)^2.
\ee
Consequently the path integral is written as
\be
Z_{\theta,\phi}[\beta]=\int \mathcal{D} \theta \mathcal{D}\phi~\exp\Big[ - \int d^2x \frac{1}{2}(\partial_\mu \theta)^2+\\
-\frac{1}{2\pi}(\partial_\mu \phi)^2 +
\frac{1}{2e^2}\partial_\mu \phi \partial^2\partial^\mu\phi - \frac{1}{2\pi}(\partial_\mu\beta)^2 \Big].
\ee
Now we perform a shift in the field $\theta$:
 \be
 \theta \to \theta+\phi/\pi+\beta/\pi
 \ee
and we integrate over the field $\phi$, 
as in the case of the pure Schwinger model. We obtain
\be
Z_{\theta}[\beta]=\int \mathcal{D}\theta e^{-\int d^2x \frac{1}{2}(\partial_\mu\theta)^2-\frac{e\partial_\mu\beta\partial_\mu\theta}{e\sqrt{\pi}}+\frac{1}{2}m^2\theta^2+\frac{1}{2}m^2\beta^2},
\ee
where $m^2=e^2/\pi$. The term proportional to $\beta^2$ can be dropped, as it does not contribute to correlation functions.  
The resulting partition function is
\be
Z_{\theta}[\beta]=\int \mathcal{D} \theta~ e^{-\int d^2x \frac{1}{2}(\partial_\mu\theta)^2+\frac{1}{2}m^2\theta^2-\frac{\partial_\mu\beta\partial_\mu\theta}{\sqrt{\pi}}}.
\ee
We note that, in such a representation,  the interaction with the external field is described  by
the term $\frac{1}{\sqrt{\pi}}\partial_\mu\beta\partial_\mu\theta.$
After re-expressing  such a coupling in terms of the original external field $A_\mu$  we find
\be
Z_{\theta}[A_\mu]=\int \mathcal{D}\theta~ e^{-\int d^2x \frac{1}{2}(\partial_\mu\theta)^2+\frac{1}{2}m^2\theta^2
-i\frac{1}{\sqrt{\pi}}A_\nu\epsilon_{\nu\mu}\partial_\mu\theta}.\nonumber\\
\ee
This result shows  that, after bosonization, the vector current operator becomes  
\be
\bar{\Psi}\gamma_\mu\Psi\rightarrow-i\frac{1}{\sqrt{\pi}}\epsilon_{\mu\nu}\partial_\nu\theta.
\ee

\end{document}